\documentclass[twocolumn,showpacs,prb]{revtex4}

\usepackage{graphicx}
\usepackage{dcolumn}
\usepackage{bm}

\usepackage{graphics}
\begin{document}

\title{Phase diagram of aggregation of oppositely
charged colloids in salty water}

\author{R. Zhang}
\author{B. I. Shklovskii}
\affiliation{Theoretical Physics Institute, University of
Minnesota, Minneapolis, Minnesota 55455}

\date{\today}

%%%%%%%%%%%%%%%%%
\begin{abstract}
%%%%%%%%%%%%%%%%%
Aggregation of two oppositely charged colloids in salty water is
studied. We focus on the role of Coulomb interaction in strongly
asymmetric systems in which the charge and size of one colloid is
much larger than the other one. In the solution, each large
colloid (macroion) attracts certain number of oppositely charged
small colloids ($Z$-ion) to form a complex. If the concentration
ratio of the two colloids is such that complexes are not strongly
charged, they condense in a macroscopic aggregate. As a result,
the phase diagram in a plane of concentrations of two colloids
consists of an aggregation domain sandwiched between two domains
of stable solutions of complexes. The aggregation domain has a
central part of total aggregation and two wings corresponding to
partial aggregation. A quantitative theory of the phase diagram in
the presence of monovalent salt is developed. It is shown that as
the Debye-H\"{u}ckel screening radius $r_s$ decreases, the
aggregation domain grows, but the relative size of the partial
aggregation domains becomes much smaller. As an important
application of the theory, we consider solutions of long
double-helix DNA with strongly charged positive spheres
(artificial chromatin). We also consider implications of our
theory for in vitro experiments with the natural chromatin.
Finally, the effect of different shapes of macroions on the phase
diagram is discussed.
\end{abstract}

\pacs{61.25.Hq, 82.70.Dd, 87.14.Gg, 87.15.Nn}

\maketitle

%%%%%%%%%%%%%%%%%%%%%%%%%%%%%%%%%%%%%%%%%%
\section{Introduction}\label{sec:introduction}
\begin{figure}[ht]
\begin{center}
\includegraphics[width=0.5\textwidth]{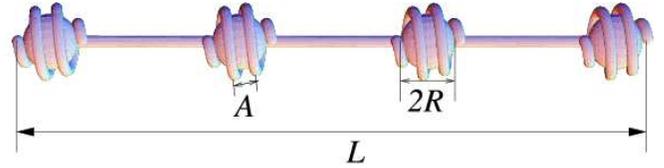}
\end{center}
\caption{A beads-on-a-string structure of the complex of a long
negative polymer macroion with positive spherical $Z$-ions
(artificial chromatin).} \label{fig:necklace}
\end{figure}

%%%%%%%%%%%%%%%%%%%%%%%%%%%%%%%%%%%%%%%%%%
Aggregation or self-assembly of oppositely charged colloids is a
general phenomenon in biology, pharmacology, and chemical
engineering. The most famous biological example is the chromatin
made of long negatively charged double-helix DNA and positively
charged histone octamers\cite{cell}. Due to Coulomb interaction,
DNA winds around many octamers to form a beads-on-a-string
structure, also called 10 nm fiber (see Fig.~\ref{fig:necklace}
for an illustration). This 10 nm fiber may self-assemble into a 30
nm fiber, which is the major building material of a chromosome.
The formation of 30 nm fiber strongly depends on the concentration
of salt in the solution. This means that the Coulomb interaction
plays a crucial role\cite{Widom}. The best known pharmacological
example is the problem of gene therapy. In this case, a negatively
charged DNA helix should penetrate through a negatively charged
cell membrane. To do this, DNA has to be neutralized or
overcharged by complexation with positive polyelectrolytes or
colloids. At the same time aggregation of these complexes can be
useful or should be
avoided\cite{Bloomfield,Sikorav,Raspaud1998,RAspaud1999,Saminathan,
Zanten2001,Zanten2002}. Industrial examples of aggregation include
using cationic polyelectrolyte as coagulants for paper
manufacturing, mineral separation, and the aggregation-induced
removal of particulate matter from the aqueous phase in water and
waste water treatment processes\cite{Walker}. Given such wide
applications, it is therefore interesting to construct a general
physical theory on aggregation of oppositely charged colloids.

\begin{figure}[ht]
\begin{center}
\includegraphics[width=0.25\textwidth]{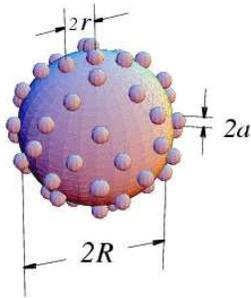}
\end{center}
\caption{A complex of a negative spherical macroion with positive
spherical $Z$-ions condensing on it.} \label{fig:sphmacroion}
\end{figure}

\begin{figure}[ht]
\begin{center}
\includegraphics[width=0.4\textwidth]{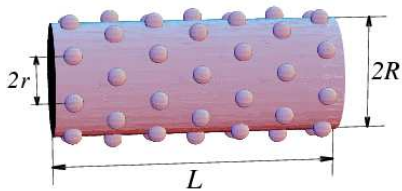}
\end{center}
\caption{A complex of a negative cylindrical macroion with
positive spherical $Z$-ions condensing on it. $Z$-ions form 2D
Wigner-crystal-like liquid on the surface of the macroion.}
\label{fig:cylinmacroion}
\end{figure}

In this paper, we consider the equilibrium state of two kinds of
oppositely charged colloids in salty water and construct the phase
diagram of such a system. Without losing generality, we call the
larger colloid ``macroion" with negative charge $-Q$,
concentration $p$, and the smaller colloid ``$Z$-ion" with
positive charge $Ze$ ($e$ is the proton charge), concentration
$s$. Macroions can be big spheres (big colloid particles), rigid
cylinders (short DNA) or long semi-flexible polymers (long DNA).
$Z$-ions can be small spheres (nucleosome core particles, micelles
or dendrimers) or short polymers (polyamines). Most of such
systems are strongly asymmetric in the sense that the size and
charge of the macroion are much larger than those of the $Z$-ion
(Figs.~\ref{fig:necklace}, \ref{fig:sphmacroion},
\ref{fig:cylinmacroion}). Therefore, defining the number of
$Z$-ions neutralizing one macroion as $N_i=Q/Ze$ (the subscript
$i$ denoting ``isoelectric"), we focus on systems with $N_i\gg1$.


\begin{figure}[ht]
\begin{center}
\includegraphics[width=0.3\textwidth]{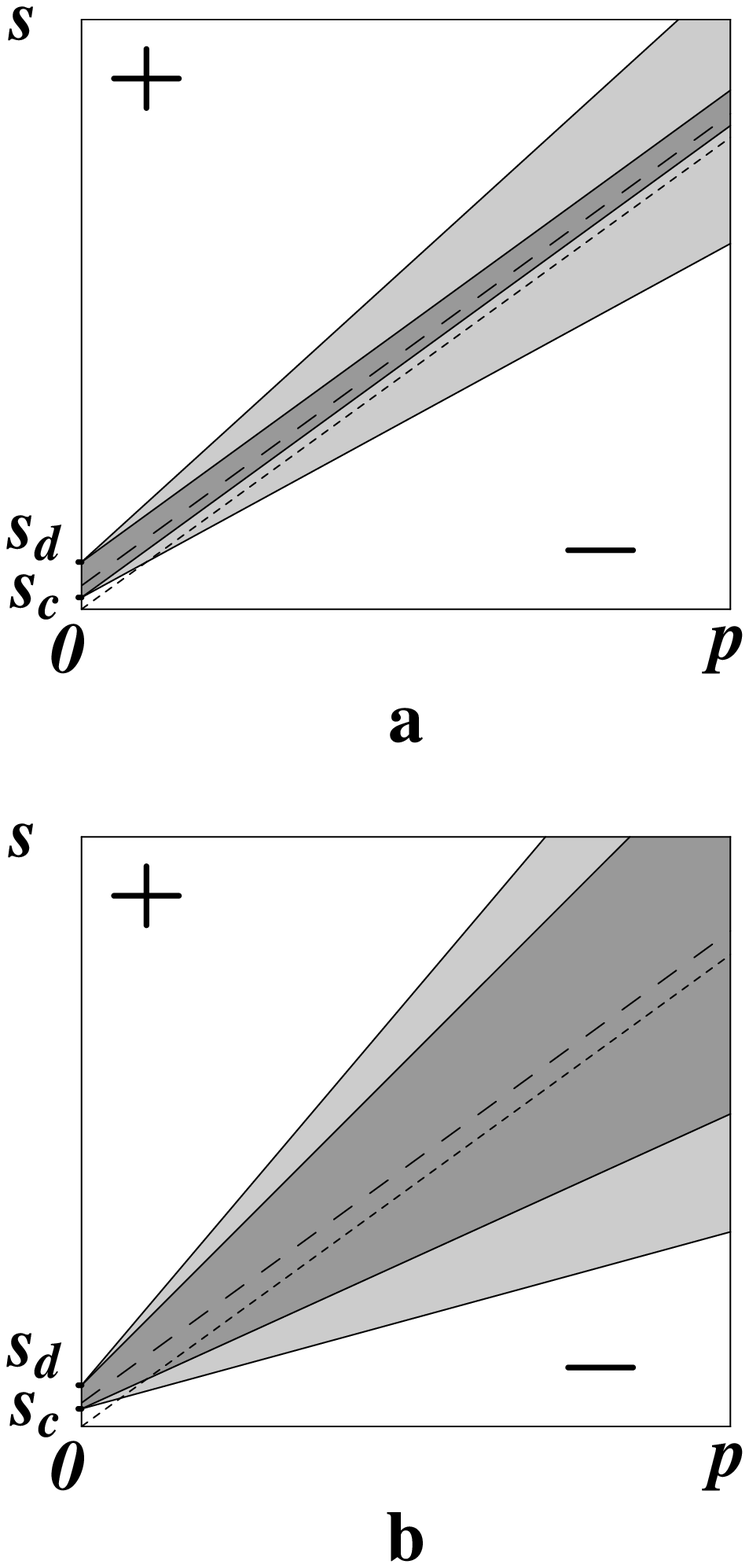}
\end{center}
\caption{Phase diagrams of the artificial chromatin system.
\textbf{a}: without monovalent salt. \textbf{b}: with monovalent
salt. (Spherical and cylindrical macroion systems have similar
phase diagrams with a much larger $s_d$.) $p$ is the concentration
of macroions (DNA), $s$ is the concentration of $Z$-ions
(spheres). Plus and minus are the signs of the charge of free
macroion-$Z$-ions complexes. The dark gray region is the domain of
total aggregation of complexes, while the light gray region is the
domain of their partial aggregation. The white region is the
domain of free complexes. $s_c$ and $s_d$ are concentrations of
$Z$-ions at the boundary of the aggregation domain when
$p\rightarrow 0$. At the dotted "isoelectric" line the absolute
values of the total charges of macroions and $Z$-ions in the
solution are equal. At the dashed "neutrality" line free complexes
are neutral.} \label{fig:basicphase}
\end{figure}

In the previous paper\cite{Nguyen-reentrant} this problem was
considered for long semi-flexible polymers (DNA double helices) as
macroions and rigid synthetic spheres with very large charge as
$Z$-ions (Fig.~\ref{fig:necklace}) without monovalent salt. In
such a system each polymer macroion winds around a number of
spherical $Z$-ions and forms a periodic necklace-like structure
which is similar to the natural chromatin. We call it ``artificial
chromatin". The phase diagram Fig.~\ref{fig:basicphase}a was
obtained in a plane of the macroion (DNA) concentration, $p$, and
the $Z$-ion (spheres) concentration, $s$ (notations $s$ and $p$
are introduced as abbreviation for ``sphere" and ``polymer"). As
seen from Fig.~\ref{fig:basicphase}a, around the line of
neutrality (the dashed line) where macroion-$Z$-ions complexes are
almost neutral, they condense in a macroscopic aggregate (the gray
region). Far away from this line, the complexes are strongly
charged and they stay free in the solution (the white region).

\begin{figure}[ht]
\begin{center}
\includegraphics[width=0.5\textwidth]{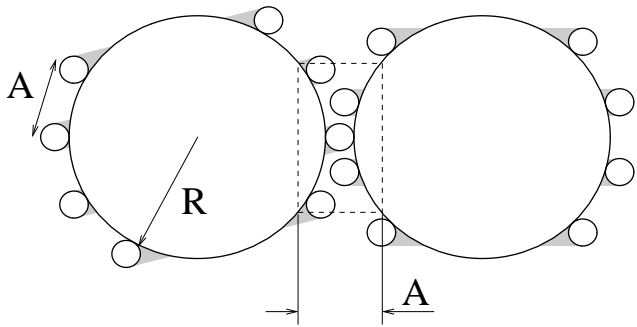}
\end{center}
\caption{Cross section through the centers of two touching
positive spherical $Z$-ions with turns of negative semi-flexible
polymers (gray) wound around them. At the place where two spheres
touch each other (the rectangular), the density of polymer
segments doubles and the correlation energy is gained.}
\label{fig:2spheres}
\end{figure}

It should be emphasized that strong correlations play a crucial
role in the origin of this picture. First, spherical $Z$-ions in
the necklace repel each other to form a 1D liquid with almost
periodic structure similar to one-dimensional Wigner crystal (see
Fig.~\ref{fig:necklace}). This leads to a correlation energy gain
in addition to the mean field Coulomb energy of the spheres. The
difference between this gain and the entropic gain in the bulk per
$Z$-ion plays the role of a voltage which may overcharge necklaces
(making them positive). As a result, a free macroion-$Z$-ions
complex can be either positive (above the dashed line in
Fig.~\ref{fig:basicphase}a) or negative (below the dashed line in
Fig.~\ref{fig:basicphase}a) depending on concentrations $s$ and
$p$. Second, the polymer turns wound around a spherical $Z$-ion
repel each other and form an almost equidistant coil. This
correlation offers a short-range attraction between complexes as
illustrated by Fig.~\ref{fig:2spheres}. When complexes are almost
neutral, this attraction is able to condense them. When they are
strongly charged, Coulomb repulsion is larger than the short-range
attraction and they stay free.

An important feature of Fig.~\ref{fig:basicphase}a is two
relatively wide domains, where aggregation of macroion-$Z$-ions
complexes is only partial (the light gray region). This means that
certain fraction of complexes aggregates, while others stay free
in the solution. The reason for partial aggregation is
redistribution of $Z$-ions between aggregated and free complexes.
When ($p$, $s$) is such that the Coulomb repulsion between
complexes is slightly larger than the short-range attraction,
$Z$-ions can redistribute themselves so that for a fraction of
complexes the Coulomb repulsion becomes substantially smaller and
they aggregate, while for the rest of complexes the Coulomb
repulsion becomes even larger and they stay free\cite{Rayleigh
note}. Without monovalent salt, the macroscopic aggregate should
be practically neutral (a net charge defined as the sum of charges
of the macroion and $Z$-ions should be close to zero). As a
result, the domain of total aggregation is narrow while the domain
of partial aggregation is much wider (see
Fig.~\ref{fig:basicphase}a).

In this paper, we develop a general phenomenological theory of
complexation and aggregation of oppositely charged macroions and
$Z$-ions in the presence of monovalent salt. In order to formulate
this theory we use the simplest model of spherical rigid macroions
and small $Z$-ions shown in Fig.~\ref{fig:sphmacroion}. In the
absence of monovalent salt, the phase diagram of this system looks
like Fig.~\ref{fig:basicphase}a, where $p$ is the concentration of
spherical macroions and $s$ is the concentration of $Z$-ions.
Screening by monovalent salt effectively truncates the Coulomb
interaction at the Debye-H\"{u}ckel screening radius $r_s$. If
$r_s$ is smaller than the size of the spherical macroions, the
Coulomb repulsion of two macroion-$Z$-ions complexes is almost
completely screened even when they touch each other. Therefore,
aggregation is possible even if macroion-$Z$-ions complexes carry
a net charge (defined as the sum of charges of the macroion and
$Z$-ions). As a result, screening strongly modifies the phase
diagram as shown in Fig.~\ref{fig:basicphase}b. Comparing with the
phase diagram without monovalent salt
(Fig.~\ref{fig:basicphase}a), we see that the aggregation domain
grows. At the same time, the relative size of the partial
aggregation domain (the light gray region) in the whole
aggregation domain decreases, because redistribution of $Z$-ions
between aggregated and free complexes becomes less important.

Our phenomenological theory of Sec. II contains several parameters
depending on charges, sizes, shapes and flexibility of macroions
and $Z$-ions. In sections III, IV and V we evaluate
microscopically these parameters for specific pairs of macroions
and $Z$-ions. In Sec. III we do this for a system of spherical
macroions and small $Z$-ions (Fig.~\ref{fig:sphmacroion}) and find
slopes of phase diagram boundaries in this case. In Sec. IV we
repeat this calculation for cylindrical macroions (Fig.~
\ref{fig:cylinmacroion}). In Sec. IV we return to artificial
chromatin (Fig.~\ref{fig:necklace}) and revise some results of
Ref.~\onlinecite{Nguyen-reentrant} related to the screening effect
of monovalent salt.

We find that although all these systems have phase diagram similar
to Fig.~\ref{fig:basicphase}b, the width of aggregation domain and
relative width of the partial aggregation domains depend on the
pair. Namely, rigid cylindrical macroions have the largest
aggregation domain, because parallel cylinders have relatively
larger area where they can touch each other. Artificial chromatin
(Fig.~\ref{fig:necklace}) on the other side has the smallest
aggregation domain. This peculiarity will be explained in Sec. V.
There we also show that phase diagram of
Fig.~\ref{fig:basicphase}b seems to agree with results of
experiments in vitro designed to understand regulation of natural
chromatin\cite{Wolffe}.

%%%%%%%%%%%%%%%%%%%%%%%%%%%%%%%%%%%%%%%%%%%%%%%%%%%%%%%%%%%%%%%%%%%%%%%%%%%%%
\section{Phenomenological theory of aggregation in the presence of monovalent salt}
\label{sec:generaltheory}
\begin{figure}[ht]
\begin{center}
\includegraphics[width=0.3\textwidth]{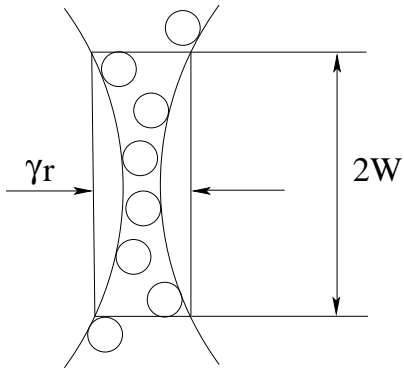}
\end{center}
\caption{The cross section of the touching area of two aggregated
spherical macroions. The sticky region is shown by the
rectangular. In this region, the density of $Z$-ions doubles and
the correlation energy is gained.}\label{fig:stickyregion}
\end{figure}

\begin{figure}[ht]
\begin{center}
\includegraphics[width=0.5\textwidth]{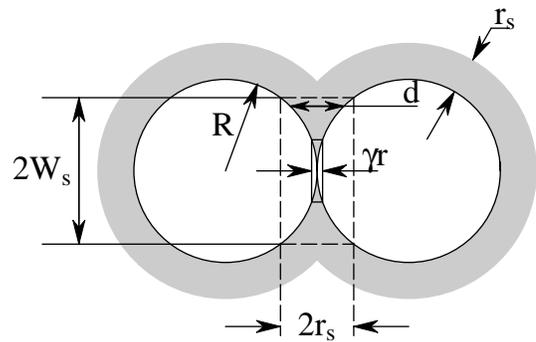}
\end{center}
\caption{The cross section of the two aggregated spherical
macroions. $Z$-ions adsorbed on them are not shown. The width of
the gray region is the effective range of the Coulomb interaction
in the presence of monovalent salt. The neutral $r_s$-cylinder is
shown by the rectangle of dashed lines. It is larger than the
sticky region shown by the full line rectangular (see
Fig.~\ref{fig:stickyregion}).} \label{fig:screeningcontact}.
\end{figure}

%%%%%%%%%%%%%%%%%%%%%%%%%%%%%%%%%%%%%%%%%%%%%%%%%%%%%%%%%%%%%%%%%%%%%%%%%%%%%
In this section we discuss the general theory of the phase diagram
in the presence of monovalent salt using the system of spherical
macroions shown in Fig.~\ref{fig:sphmacroion}. Aggregation of two
such spherical macroions is illustrated by
Figs.~\ref{fig:stickyregion} and ~\ref{fig:screeningcontact}.
$Z$-ions form a strong correlated liquid on the surface of the
macroion. The negative correlation energy in this liquid results
in the voltage which may overcharge the complex. It also induces a
short-range attraction between complexes since in the spot where
two complexes touch each other (see Fig.~\ref{fig:stickyregion})
the surface density of the correlated liquid is doubled and the
energy per $Z$-ion is reduced~\cite{BR}. This attractive force
decays as $\exp( -2\pi d/\sqrt{3}r)$ (for a triangular
lattice)\cite{Ruzin} with the distance $d$ between two surfaces,
where $r$ is the half of average distance between nearest neighbor
$Z$-ions on the macroion surface, or, in other words, the radius
of the Wigner-Seitz cell. Consequently, only a small disk-like
contact region with radius $W = \sqrt{\gamma Rr}$, where $\gamma
=\sqrt{3}/2\pi\sim 0.3$, contributes to the attraction of two
complexes (see Fig.~\ref{fig:stickyregion}). We call it ``sticky
region".

Similarly to artificial chromatin, this correlation attraction
leads to phase diagram Fig.~\ref{fig:basicphase}a in the absence
of screening. We should remember, however, that in the case of
Fig.~\ref{fig:sphmacroion}, $s$ stands for the concentration of
small spherical $Z$-ions, and $p$ is the concentration of large
spherical macroions. We show below that screening not only
enlarges the aggregation domains, but also leads to relatively
smaller size of the partial aggregation domain. These effects
become strong when $r_s$ is smaller than $R$. They grow when $r_s$
is reduced all the way to $r$ and even at $r_s<r$. Only when $r_s$
is so small that interaction of a $Z$-ion with macroion surface
becomes smaller than $k_BT$, most of $Z$-ions leave the macroion
surface, and aggregation becomes impossible. Below we concentrate
on the case $r\ll r_s\ll R$ which can be described by simple and
universal phenomenological theory.

There is a key difference between the cases without screening
discussed in Ref.~\onlinecite{Nguyen-reentrant} and with screening
effect of monovalent salt. In the first case, each macroion of the
aggregate can only carry $N_i$ $Z$-ions, i.e., the aggregate is
almost neutral. This is because if macroion-$Z$-ions complexes
were charged, the Coulomb potential of the aggregate would
increase with its size and become too large to continue
aggregation. On the other hand, in the presence of strong
screening by monovalent salt, when the local distance between the
two touching surfaces, $d$, is much larger than $r_s$, the Coulomb
interaction between two surfaces is exponentially small and can be
neglected (see Fig.~\ref{fig:screeningcontact}). It only affects
the disk-like surface region with radius $W_s=\sqrt{2Rr_s}$
located inside the cylinder where $d<2r_s$. We refer to such a
cylinder as ``$r_s$-cylinder" (shown by the dashed rectangular in
Fig.~\ref{fig:screeningcontact}). Therefore, if we count only
charges of macroions and $Z$-ions, the aggregate of
macroion-$Z$-ions complexes need not be neutral. Most of the
surface area of these complexes can be overcharged or undercharged
in the same way as surfaces of free complexes.

We argue that macroion surface charge density, $\sigma(d)$
(including charge of the macroion and Z-ions), inside
$r_s$-cylinder is much smaller than the one outside
$r_s$-cylinder, $\sigma_0$. In electrostatics, it is known that
for two charged contacting metallic spheres in vacuum, the surface
charge density goes to zero when we move to the point of contact
(one can show that it decays as $\exp(-\pi\sqrt{R/d})$). Our case
is similar to this example in the sense that $Z$-ions equilibrate
on macroion surfaces so that everywhere they acquire the same
electrochemical potential, $\mu_{c}(N)+ e\phi$. The chemical
potential $\mu_{c}(N)$ determined by correlations of $Z$-ions
depends on concentration of $Z$-ions on the surface of macroion,
which, as shown below, can be regarded as a constant. This means
that the electrostatic potential $\phi$ is a constant, too,
similarly to the case of two metallic spheres. The peculiarity of
our case is that the presence of monovalent salt which screens the
surface electric potential. Deep inside $r_s$-cylinder, screening
by monovalent salt is incomplete. Indeed, according to the
Debye-H\"{u}ckel theory, the density of screening charge of
monovalent salt near the macroion surface has the same magnitude
$\rho = -\phi/4\pi r_{s}^{2}$ as outside of $r_s$-cylinder. But
with decreasing $d$, the volume occupied by this charge shrinks
very fast. At $d\ll r_s$, charge density of monovalent salt
projected to the macroion surface, $\sigma^{\prime}=-\phi d/4\pi
r_s^2$, is much smaller than $\sigma_0$. As we saw from the
problem of two contacting metallic spheres, the total surface
charge density between two touching spherical surfaces with
constant potential decays very fast towards the contact point.
Accordingly, inside the $r_s$-cylinder at $d\ll r_s$, the surface
charge density $\sigma+\sigma^{\prime}/2$ is very small (it decays
with decreasing $d$ as
$\sigma_0\exp(\pi(\sqrt{R/2r_s}-\sqrt{R/d}))$). Therefore, for the
surface charge density of the macroion $\sigma(d)$, we arrive at
$\sigma(d)=-\sigma^{\prime}(d)/2=\phi d/8\pi r_{s}^{2}$. Since at
$d \ll r_s$, $\sigma(d)$ is much smaller than $\sigma_0$, we call
this region ``neutral region". At $r\ll r_s$, the sticky region
discussed above (the full line rectangular in
Figs.~\ref{fig:stickyregion} and \ref{fig:screeningcontact}) is
much smaller than the neutral region and therefore almost neutral.

Let $N$ be the number of $Z$-ions in a free macroion-$Z$-ions
complex. We define $Q^{\ast}=-Q + NZe= Ze(N-N_i)$ as the net
charge of a free complex, and $(1-\beta) Q^{\ast}$ as the net
charge of the aggregated complex. The total number of $Z$-ions in
an aggregated complex is therefore $N-\beta(N-N_i)$.

At $r_s\gg R$, the whole complex is neutral in the aggregate, so
$\beta=1$. At $r_s\ll R$, away from neutral regions, due to
equilibration of Z-ions, aggregated macroion-$Z$-ions complexes
have the same surface density of $Z$-ions as free complexes. Thus,
$\beta$ is a measure of the fraction of total surface area of a
complex occupied by neutral regions. $\beta$ decreases with
decreasing $r_s$.

Now we can write down the free energy of the system. Similarly to
Ref.~\onlinecite{Nguyen-reentrant}, we start from a point $(p,s)$
in the domain of partial aggregation and assume that a fraction
$x$ of aggregated macroion-$Z$-ions complexes. The free energy per
unit volume is

\begin{eqnarray}
&F&(N,x;s,p)=(1-x)p\left[\frac{Q^{\ast2}}{2C}+NE(N)\right]\nonumber\\
&+&xp\left[(1-\beta)\frac{Q^{\ast2}}{2C} + NE(N)- \beta \mu_c
(N-N_i)+\epsilon\right]
\nonumber\\
&+&k_BT[s-(1-x)pN-xp(N-\beta(N-N_i))]\nonumber\\
&\times&\ln\frac{[s-(1-x)pN-xp(N-\beta(N-N_i))]v_0}{e}.\label{screeningfree}
\end{eqnarray}
The first line of Eq.~(\ref{screeningfree}) is the free energy
density of free macroion-$Z$-ions complexes. It consists of
Coulomb energy of complexes and correlation energy of $Z$-ions.
Here $C$ is the capacitance of a free complex, $E(N)<0$ is the
correlation energy per $Z$-ion in the complex. The second line of
Eq.~(\ref{screeningfree}) is the free energy density of aggregated
complexes. Here $\epsilon<0$ is the additional correlation energy
per complex due to aggregation\cite{notation} and
$\mu_c(N)=\partial(NE(N))/\partial N<0$ is the part of the
chemical potential of $Z$-ions related to their correlations on
the surface of a macroion. It has been assumed in
Eq.~(\ref{screeningfree}) that $N-N_i\ll N_i$ (this will be
confirmed later) and therefore
\begin{equation}
\mu_c(N)=\mu_c(N_i)\equiv\mu_c.\label{mucNNi}
\end{equation}
In the second line of Eq.~(\ref{screeningfree}) the first term is
the Coulomb energy of an aggregated complex diminished by a loss
of fraction $\beta$ of capacitor area, the second and third term
give correlation energy of $Z$-ions diminished by loss of $\beta
(N-N_i)$ of them and the fourth term is responsible for attraction
energy of aggregated complexes in sticky regions. The third and
fourth line  of Eq.~(\ref{screeningfree}) give the free energy
density of free $Z$-ions related to their entropy, ($v_0$ is the
normalizing volume). The entropy of macroions is ignored because
of their much smaller concentration. Minimizing this free energy
with respect to $N$ and $x$, we get
\begin{eqnarray}
\mu_c+\frac{ZeQ^{\ast}}{C}&=&k_BT\ln[(s-(1-x)pN\nonumber\\
&-&xp(N-\beta(N-N_i)))v_0],
\label{mucscreeneq}\\
\epsilon-\beta\frac{Q^{\ast2}}{2C}&=&
\beta(N-N_i)\{\mu_c-k_BT\ln[(s-(1-x)pN\nonumber\\
&-&xp(N-\beta(N-N_i)))v_0]\},\label{epsilonscreeneq}
\end{eqnarray}

Excluding the entropy term from in Eqs.~(\ref{mucscreeneq}) and
(\ref{epsilonscreeneq}), we get
\begin{equation}
N_{c,d}=N_i\left(1\mp\sqrt{\frac{|\epsilon|}{\beta(r_s)}\frac{2C(r_s)}{Q^2}}\right),
\label{Ncdscreen}
\end{equation}
Here and below, the upper (lower) sign in the formula always
corresponds to the first (second) subscript of the symbol.

Plugging Eq.~(\ref{Ncdscreen}) back to Eq.~(\ref{mucscreeneq}), we
get
\begin{equation}
s_{c,d}(p)=s_{c,d}+(1-x)pN_{c,d}+xp[N_{c,d}-\beta(r_s)(N_{c,d}-
N_i)],\label{scdp}
\end{equation}
where
\begin{equation}
 s_{c,d}=\frac{1}{v_0}\exp\left[\frac{1}{k_BT}
\left(\mp\sqrt{\frac{|\epsilon|}{\beta(r_s)}\frac{2Z^2e^2}{C(r_s)}}
-|\mu_c|\right)\right].\label{scdscreen}\nonumber\\
\end{equation}

The two solutions $N_{c,d}$ and $s_{c,d}$ mean that we have two
partial aggregation domains. Taking $x=0$ and $x=1$ in
Eq.~(\ref{scdp}), we get two outer boundaries
\begin{equation}
s_{c,d}(p;x=0)=s_{c,d}+pN_{c,d},\label{scdx0}
\end{equation}
and two inner boundaries
\begin{equation}
s_{c,d}(p;x=1)=s_{c,d}+p[N_{c,d}-\beta(r_s)(N_{c,d}-N_i)]\label{scdx1}
\end{equation}
of the partial aggregation domains. The corresponding phase
diagram is shown in Fig.~\ref{fig:basicphase}b. The outer
boundaries are also the boundaries of the whole aggregation domain
(including partial and total aggregation domains), at which all
free macroion-$Z$-ions complexes are still stable ($x=0$). The
inner boundaries are also the boundaries of the total aggregation
domain at which there is no free macroion-$Z$-ions complex
($x=1$). When $s$ increases from zero, first the aggregate forms,
then it dissolves. Thus we arrive at a phase diagram of reentrant
condensation. According to Eq.~(\ref{scdx0}), the slopes, $ds/dp$,
of the outer boundaries are $N_{c,d}$, which are the numbers of
$Z$-ions in a free macroion-$Z$-ions complex. Since $N_{c,d}$ are
solutions to $N$ in the whole partial aggregation domain, the
number of $Z$-ions complexed with one free macroion is fixed in
the partial aggregation domain. According to Eq.~(\ref{scdx1}),
the slopes of the inner boundaries are
$N_{c,d}-\beta(N_{c,d}-N_i)$, which are the numbers of $Z$-ions in
an aggregated complex. On the other hand, $s_{c,d}$ are the
intercepts of these boundaries with $p$ axis. The dashed line on
Fig.~\ref{fig:basicphase}b corresponds to solutions with neutral
macroion-$Z$-ions complexes and can be calculated from
Eq.~(\ref{mucscreeneq}). Taking $N=N_i$ in this equation, we get
\begin{equation}
s_0(p)=\frac{1}{v_0}\exp\left(-\frac{|\mu_c|}{k_BT}\right)+pN_i.\label{dashline}
\end{equation}
This line has the same slope as the isoelectric line $s_i=pN_i$
(dotted line in Fig.~\ref{fig:basicphase}), but a small finite
intercept. This is because there is always a small fraction of
free $Z$-ions in the solution.

When $\beta=1$, the whole surface of a macroion in the aggregate
is neutral, all formulae above reproduce the corresponding results
in the case $r_s\gg R$\cite{Nguyen-reentrant}. For example, the
boundaries of the total aggregation domain given by
Eq.~(\ref{scdx1}) are now parallel lines
\begin{equation}
s_{c,d}(p;x=1)=s_{c,d}+pN_i.\label{scdx1Ni}
\end{equation}
The corresponding phase diagram for $r_s\gg R$ is shown in
Fig.~\ref{fig:basicphase}a.

There are four parameters $\mu_c$, $\epsilon$, $C$ and $\beta$ in
the theory. Their values depend on the screening radius $r_s$, the
shape and flexibility of the macroions and $Z$-ions. In the
following sections, we calculate them for specific systems. But
even now we can qualitatively summarize the evolution of the phase
diagram with decreasing $r_s$. For this purpose, we define the
size of an aggregation domain as the absolute value of the
difference of the slopes of its two boundaries. As $r_s$
decreases, $\beta(r_s)$ decreases, $C(r_s)$ increases, while
$\epsilon$ is fixed (this will be shown below). Then according to
Eq.~(\ref{Ncdscreen}), $N_d$ increases and $N_c$ decreases. Thus,
both the size of the whole aggregation domain (including partial
and total aggregation domains), $N_d-N_c$, and the size of the
total aggregation domain, $(1-\beta)(N_d-N_c)$, grow with
decreasing $r_s$. In contrary to them, the relative size of the
two partial aggregation domains in the whole aggregation domain,
$\beta(N_d-N_c)/(N_d-N_c)=\beta$, decreases with decreasing $r_s$.

As we mentioned before, at $r_s<r$ the growth of the aggregation
domain and reduction of the relative size of the partial
aggregation domain continue with decreasing $r_s$. In this regime,
however, the range of Coulomb interaction and correlation becomes
the same and one can not separate these interactions from each
other. A microscopic theory is necessary, which operates with
screened interaction of $Z$-ions with macroion and with each other
and allows for small dielectric constant of macroion. This is a
subject of future work.

%%%%%%%%%%%%%%%%%%%%%%%%%%%%%%%%%%%%%%%%%%%%%%%%%%%%%%%%%%%
\section{Rigid spherical macroions}\label{sec:sphere}
%%%%%%%%%%%%%%%%%%%%%%%%%%%%%%%%%%%%%%%%%%%%%%%%%%%%%%%%%%%
In this section, we consider the system with spherical macroions
and much smaller spherical $Z$-ions (Figs.~\ref{fig:sphmacroion}).
We estimate parameters $\mu_c$, $\epsilon$, $C$ and $\beta$
microscopically and find out how borders of the aggregation domain
change with decreasing screening radius $r_s$.

Let us first consider the case $r_s\gg R$. In this case, each
macroion carry $N_i$ $Z$-ions in the aggregate and $\beta=1$. In
the first order approximation, $Z$-ions repel each other to form
2D Wigner crystal on the surface of a macroion. Thus the
correlation chemical potential $\mu_c$ is
approximately\cite{Nguyen-review}
\begin{equation}
\mu_c=-\frac{1.6Z^2e^2}{Dr},\label{muc}
\end{equation}
where $r$ is the radius of the Wigner-Seitz cell when a macroion
is neutralized by $Z$-ions (see Figs.~\ref{fig:sphmacroion}) and
$D$ is the dielectric constant of water. To calculate $\epsilon$,
we first notice that the number of $Z$-ions in a sticky region is
\begin{equation}
N_i\frac{\pi W^2}{4\pi R^2}=\frac{\gamma\sqrt{N_i}}{2},
\end{equation}
where relation $N_i=4\pi R^2/\pi r^2$ has been used. If each
macroion has $b$ nearest neighbors in the aggregate ($b=12$ for
dense packing),
\begin{equation}
\epsilon=\frac{b\gamma}{2}\sqrt{N_i}\alpha\mu_c=-0.4b\gamma\alpha\frac{ZeQ}{DR}.
\label{epsilon}
\end{equation}
Here we assumed that each $Z$-ion in the sticky region gains the
correlation energy $\alpha\mu_c$. The maximum possible $\alpha$
can be estimated as follows. Since the surface density of $Z$-ions
is doubled in the contact region, the radius of the Wigner-Seitz
cell becomes $r/\sqrt{2}$ and the correlation energy per $Z$-ion
gets a factor $\sqrt{2}$. This gives $\alpha\simeq 0.3$. The real
$\alpha$ is much smaller than this upper limit due to the effect
of small dielectric constant of macroions. As is well known in
electrostatics, positive $Z$-ions in water ($D\simeq 80$) creates
positive images in adjacent macroions ($D\ll 80$). Repulsion from
images push two contacting macroions away from each other and
diminishes the gain of correlational energy. Finally, for a
spherical macroion with $Z$-ions whose size is negligible, the
capacitance is
\begin{equation}
C=DR.\label{spherecapacitance}
\end{equation}
Using Eqs.~(\ref{Ncdscreen}) and (\ref{scdscreen}), we get
\begin{eqnarray}
N_{c,d}&=&N_i\left(1\mp\sqrt{0.2b\gamma\alpha}\frac{r}{R}\right),\label{Ncdsphere}\\
s_{c,d}&=&\frac{1}{v_0}\exp\left[-\frac{1.6Z^2e^2}{k_BTDr}
\left(1\pm\sqrt{\frac{b\gamma\alpha}{0.8}}\right)\right].\label{scdsphere}
\end{eqnarray}

Let us now switch to the case $r\ll r_s\ll R$. Since the sticky
region is much smaller than the neutral region (see
Fig.~\ref{fig:screeningcontact}), the short-range correlations are
not changed by screening, $\mu_c$ and $\epsilon$ are still given
by Eqs.~(\ref{muc}) and (\ref{epsilon}). But $\beta$ and $C$ are
affected by screening. For simplicity, we estimate $\beta$ by
assuming that the whole $r_s$-cylinder is neutral. Then $\beta$
reduces to the area ratio of the neutral region to the whole
surface of the complex,

\begin{equation}
\beta=b\frac{\pi W_s^2}{4\pi R^2}=\frac{br_s}{2R}. \label{beta}
\end{equation}
And
\begin{equation}
C=\frac{DR^2}{r_s}.\label{spherecscreen}
\end{equation}
This gives
\begin{eqnarray}
N_{c,d}&=&N_i\left(1\mp\sqrt{0.4\gamma\alpha}\frac{r}{r_s}\right),
\label{Ncdspherescreen}\\
s_{c,d}&=&\frac{1}{v_0}\exp\left[-\frac{1.6Z^2e^2}{k_BTDr}
\left(1\pm\sqrt{\frac{\gamma\alpha}{0.4}}\right)\right].
\label{scdscreensphere}
\end{eqnarray}
It is easy to check that in both cases $N_{c,d}$ are close to
$N_i$. This justifies the approximation~(\ref{mucNNi}) used in the
last section. From Eq.~(\ref{Ncdspherescreen}), we see that the
aggregation domain broadens with decreasing $r_s$ and the angle it
occupies becomes the order of unity at $ r_s \sim r$. According to
Eq.~(\ref{beta}), the relative size of the partial aggregation
domain decreases with $r_s$ proportionally to $r_s/R$.

Two comments on validity of our approach are in order here. First,
we assumed above that at $s > s_{d}(p)$, when overcharged
macroions redissolve, they still keep almost all $Z$-ions. As is
well known\cite{Nguyen-review}, the condition of maximum charge
inversion is given by
\begin{equation}
\mu_c=\frac{ZeQ^{\ast}}{C}.
\end{equation}
Denoting $N_{max}$ the maximum number of Z-ions on a macroion, and
using Eq.~(\ref{muc}), in the case of $r_s \ll R$ we have
\begin{equation}
N_0=N_i\left(1+0.4\frac{r}{r_s}\right)
\end{equation}
Comparing with Eq. (\ref{Ncdspherescreen}), we see that $N_{max}$
and $N_d$ are of the same order of magnitude. As we mentioned
above we believe that $\alpha < 0.3$ so that $N_d < N_{max}$ and
the picture of stable complexes on both sides of aggregation
domain is justified.

Second, up to now we discussed screening by monovalent salt in
Debye-H\"{u}ckel approximation. When a $Z$-ion has large charge
and small size $a$, they are screened nonlinearly. Monovalent
counterions condense on $Z$-ions (similarly to the Manning
screening of a charged cylinder) and renormalize the charge of a
free $Z$-ion to a smaller value
$Z^{\prime}$\cite{Nonlinearscreen}. The charge $Z^{\prime}$ is
screened in Debye-H\"{u}ckel way. When adsorbed to the macroion
surface, $Z$-ion is subjected to the additional Coulomb
interactions with both the negative macroion surface and
neighboring $Z$-ions. Both interaction energies are of the order
of $Z^{\prime}e^2/r$ and at $a \ll r$ can be ignored in comparison
with with the chemical potential of the monovalent counterion
$Z^{\prime}e^2/a$ on $Z$-ion. Therefore, the effective charge
$Z^{\prime}$ is not changed by $Z$-ion condensation and should be
used as the effective charge of a $Z$-ion in all our
results\cite{Nguyen-salty}.

In recent Monte Carlo simulations of the system of spherical
macroions with multivalent $Z$-ions\cite{Lobaskin},  it is seen
that around the isoelectric line $s_i=pN_i$, all macroions
aggregate, while far away from this line on both sides, free
macroion-$Z$-ions complexes are stable. This qualitatively agrees
with our theory.

%%%%%%%%%%%%%%%%%%%%%%%%%%%%%%%%%%%%%%%%%%%%%%%%%%%%%%%%%%%%%%%%%
\section{Cylindrical macroions}\label{sec:polymer}
%%%%%%%%%%%%%%%%%%%%%%%%%%%%%%%%%%%%%%%%%%%%%%%%%%%%%%%%%%%%%

In this section, we consider the system with cylindrical macroions
and much smaller spherical $Z$-ions
(Fig.~\ref{fig:cylinmacroion}). We show that the qualitative
feature of the phase diagram of this system is the same as
spherical macroions. The major difference is that for cylinders,
$s_d$ is much larger.

We assume that charge of the macroion and $Z$-ion is such that
$r\ll R$. Accordingly two dimensional Wigner-Crystal-like
structure of $Z$-ions can be formed on the surface of
macroions\cite{Shklov99}. We also assume that $r_s<L$.
Consequently below we focus on two cases corresponding to $R\ll
r_s\ll L$ and $r\ll r_s\ll R$. Also note that in order to gain
more contact area and, therefore, more correlation energy, all
cylinders in the aggregate are parallel to each other.

We first consider the case $R\ll r_s\ll L$. It is easy to see that
the aggregate should be neutral, i.e., $\beta=1$, and the chemical
potential $\mu_c$ is still given by Eq.~(\ref{muc}). However,
$\epsilon$ is not the same since the area of the sticky region is
much larger. It is not a disk, but a stripe with area $2WL$. The
number of $Z$-ions in the sticky region is
\begin{equation}
N_i\frac{2WL}{2\pi RL}=\frac{1}{\pi}N_i\sqrt{\frac{\gamma r}{R}}.
\end{equation}
For a macroion with $b$ nearest neighbors ($b=6$ for dense
packing), the additional correlation energy per macroion is
\begin{equation}
\epsilon=\frac{b\alpha}{\pi}N_i\mu_c\sqrt{\frac{\gamma
r}{R}}.\label{epsiloncylinder}
\end{equation}
Here $\alpha$ takes the same value as the one for spherical
macroions. And the capacitance is
\begin{equation}
C=\frac{DL}{2\ln(r_s/R)}.
\end{equation}
Using Eqs.~(\ref{Ncdscreen}) and (\ref{scdscreen}), we have
\begin{eqnarray}
N_{c,d}&=&N_i\left[1\mp\sqrt{\frac{0.8b\alpha\gamma^{\frac{1}{2}}}{\pi\ln(r_s/R)}}\left(
\frac{r}{R}\right)^{\frac{3}{4}}\right],
\label{Ncdcylinder}\\
s_{c,d}&=&\frac{1}{v_0}\exp\left[-\frac{1.6Z^2e^2}{k_BTDr}
\left(1\pm\sqrt{\frac{b\alpha\gamma^{\frac{1}{2}}}{0.2\pi}\ln\frac{r_s}{R}}
\left(\frac{R}{r}\right)^{\frac{1}{4}}\right)\right].\label{scdcylinder}\nonumber\\
\end{eqnarray}

Now we consider the case $r\ll r_s\ll R$. Instead of a neutral
aggregate, we only require a stripe-like neutral region in the
present case. It is easy to see that both $\mu_c$ and $\epsilon$
are not changed by screening. The fraction of the neutral area is
\begin{equation}
\beta=b\frac{2W_sL}{2\pi
RL}=\frac{b}{\pi}\sqrt{\frac{2r_s}{R}},\label{betacylin}
\end{equation}
while the capacitance
\begin{equation}
C=\frac{DRL}{2r_s}.
\end{equation}
We have
\begin{eqnarray}
N_{c,d}&=&N_i\left[1\mp\sqrt{0.8\alpha\left(\frac{\gamma}{2}\right)^{\frac{1}{2}}}
\left(\frac{r}{r_s}\right)^{\frac{3}{4}}\right],
\label{Ncdcylinderscreen}\\
s_{c,d}&=&\frac{1}{v_0}\exp\left[-\frac{1.6Z^2e^2}{k_BTDr}
\left(1\pm\sqrt{\frac{\alpha}{0.2}\left(\frac{\gamma}{2}\right)^{\frac{1}{2}}}
\left(\frac{r_s}{r}\right)^{\frac{1}{4}}\right)\right].
\label{scdcylinderscreen}\nonumber\\
\end{eqnarray}

Again in both cases $N_{c,d}$ are close to $N_i$ and the
approximation~(\ref{mucNNi}) is valid. According to
Eq.~(\ref{betacylin}), the relative size of the partial
aggregation domain decreases with $r_s$ proportionally to
$\sqrt{r_s/R}$. We see that the aggregation domain are wider in
the case of cylindrical macroions than that for spherical
macroions (see Sec. III). This happens because of the larger
contact area in the former case.

According to Eq.~(\ref{scdcylinderscreen}), formally $s_d$ can be
even larger than $1/v_0$. Consequently, one may conclude that the
aggregate never dissolve at large $s$. Actually, this is not true.
When $s$ becomes so large that the distance between $Z$-ions in
the bulk is equal to $r_s$, the repulsion between free $Z$-ions
can not be ignored and our theory is invalid. At such a large
concentration, $Z$-ions are correlated not only on the surface of
macroions but also in the bulk and aggregation does not happen.
What we can conclude is that for systems discussed in this
section, the aggregate dissolves at very large $s$.
Mathematically, $s_d$ is so large that
$\ln(1/s_dv_0)\leq\ln(s_d/s_c)$. Correspondingly, the aggregation
domain in Fig.~\ref{fig:basicphase} at $s>s_i$ is very wide.

The results above can be qualitatively compared with experimental
results for solutions of DNA with spermine
~\cite{Sikorav,Raspaud1998,RAspaud1999,Saminathan} in which long
DNA double helices may be considered as rigid negative cylinders
and short positive spermine molecules as small $Z$-ions ($Z= 4$).
In experiments, reentrant condensation is observed. Also the
aggregation domain at small $p$ is very wide ($\log(s_d/s_c)\simeq
4$). Both facts agree with our theory.

%%%%%%%%%%%%%%%%%%%%%%%%%%%%%%%%%%%%%%%%%%%%%%%%%%%%%%%%%%%%%%%%%
\section{artificial chromatin}\label{sec:polymer}
%%%%%%%%%%%%%%%%%%%%%%%%%%%%%%%%%%%%%%%%%%%%%%%%%%%%%%%%%%%%%

In this section, we discuss the artificial chromatin in which
macroions are long semi-flexible polymers with linear charge
density $-\eta$ and $Z$-ions are hard spheres with radius $R$
(Fig.~\ref{fig:necklace}). We assume $R\eta\ll Ze$, i.e., many
turns of polymer segments are needed to neutralize one sphere. The
aggregate of such a system looks like Fig.~3 in
Ref.~\onlinecite{Nguyen-reentrant}. The phase diagram of this
system has already been discussed in detail in
Ref.~\onlinecite{Nguyen-reentrant} under the assumption that the
aggregate is always neutral which is valid only for $r_s\gg R$.
When $r_s\ll R$, there is no justification for the neutrality of
the aggregate and the theory in Sec.~\ref{sec:generaltheory}
should be used. Here we calculate the parameters microscopically
and obtain the phase diagram for the later case.

First, let us remind the results following directly from
Ref.~\onlinecite{Nguyen-reentrant}. In this case, each
macroion-$Z$-ions complex carries $N_i$ $Z$-ions in the aggregate.
Correspondingly, $\beta=1$. The correlation chemical potential
$\mu_c$ is essentially the self-energy of a bare free sphere in
the solution which is almost totally eliminated in the complex.
Therefore
\begin{equation}
\mu_c=-\frac{Z^2e^2}{2DR},
\end{equation}
where $D$ is the dielectric constant in water solution. To
calculate $\epsilon$, we first notice that the correlation energy
per unit length of the polymer is just the interaction energy of
the polymer turn with its stripe of background (sphere) positive
charge. Thus, it is $-\eta^2\ln(R/A)/D$. Correspondingly, when the
density doubles, $A$ is halved, the gain in correlation energy per
unit length is $-\alpha\eta^2/D$, where $\alpha=\ln2$ as a rough
estimate. Similar to the case of spherical macroion, the radius of
the sticky region $W\simeq\sqrt{RA/2}$. And the polymer length in
this region is $\pi W^2/A=\pi R/2$. Since each aggregated complex
carries $N_i$ $Z$-ions and has $b$ nearest neighbors, we have
\begin{equation}
\epsilon=-\frac{\pi
b\alpha}{2}\frac{R\eta^2}{D}N_i,\label{epsilonpolymer}
\end{equation}
where $b=6$ if the maximum packing number is achieved. The
capacitance
\begin{equation}
C=\frac{DL}{2\ln(r_s/R)}=\frac{DRN_i}{\ln(r_s/R)},
\end{equation}
where we have used $L\simeq 2RN_i$, because near neutrality, each
polymer absorbs about $N_i$ spheres and all spheres in the
macroion-$Z$-ions complex are densely packed. Using
Eqs.~(\ref{Ncdscreen}) and (\ref{scdscreen}), we have
\begin{eqnarray}
N_{c,d}&=&N_i\left(1\mp\sqrt{\frac{\pi
b\alpha}{\ln(r_s/R)}}\frac{R\eta}{Ze}\right),
\label{Ncdpolymer}\\
s_{c,d}&=&\frac{1}{v_0}\exp\left[-\frac{Z^2e^2}{2k_BTDR}
\left(1\pm2\sqrt{\pi
b\alpha\ln\frac{r_s}{R}}\frac{R\eta}{Ze}\right)\right].
\label{scdpolymer}\nonumber\\
\end{eqnarray}
This gives a typical phase diagram of reentrant condensation
(Fig.~\ref{fig:basicphase}a).

\begin{figure}[ht]
\begin{center}
\includegraphics[width=0.3\textwidth]{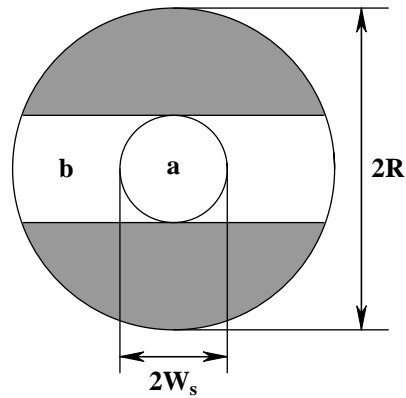}
\end{center}
\caption{A 2D view of the surface of a sphere touching the other
sphere at the center of the disk-like region \textbf{a}. To make
\textbf{a} neutral, a semi-flexible polymer have to change the
density of its turns to make the whole equatorial stripe-like
region \textbf{b} neutral. The rest of the surface colored in gray
is charged.} \label{fig:spot}
\end{figure}

Now let us consider the case when $A\ll r_s\ll R$. $A$ being the
average distance between two polymer turns on the surface of
spheres when polymer segments exactly neutralize the sphere (see
Fig.~\ref{fig:necklace}). Since $\eta/A\simeq Ze/R^2$, the
requirement $A\ll R$ is equivalent to $R\eta\ll Ze$. As we
discussed in Sec.~\ref{sec:generaltheory}, with the screening of
monovalent salt, the whole aggregate need not be neutral, but
disk-like regions on spheres where two spheres touch each other
must be neutral (see Fig.~\ref{fig:screeningcontact}). The radius
of this region $W_s=\sqrt{2r_sR}$. In the present case, the
neutral region should be made by the rearrangement of the polymer
segments on the surface of spheres. If the polymer is extremely
flexible, it can bend easily in the disk-like region. By this
bending, the surface density of polymer segments is changed
``locally" so that the region can be made neutral and the rest of
the surface is not changed. However, if the polymer is only
semi-flexible (the persistent length $l_p$ is such that
$\sqrt{r_sR}\ll l_p\ll R$), it is too rigid to neutralize the
disk-like region. Then this region can be made neutral only by
changing distances between sequential turns of the polymer on the
surface of the sphere. For example, for two touching spheres
(Fig.~\ref{fig:2spheres}), the equatorial stripe-like region shown
in Fig.~\ref{fig:spot} can be made neutral. We focus on the case
of the semi-flexible polymer which corresponds to DNA.

When $A\ll r_s\ll R$, the correlation chemical potential $\mu_c$
is still given by the self-energy of a bare free sphere, which is
now
\begin{equation}
\mu_c=-\frac{r_sZ^2e^2}{2DR^2},\label{mucpolymer}
\end{equation}
and the correlation gain due to aggregation, $\epsilon$, is still
given by Eq.~(\ref{epsilonpolymer}). This is because the contact
region where the additional correlational energy is gained is much
smaller than the neutral region as discussed in
Sec.~\ref{sec:generaltheory}. Also since the number of $Z$-ions in
each complex in the aggregate is close to $N_i$, in the first
order approximation, we can use $N_i$ as the number of $Z$-ions in
each complexes in Eq.~(\ref{epsilonpolymer}). Knowing that the
area of the strip-like region is $4\pi W_sR$ (see
Fig.~\ref{fig:spot}), the area ratio $\beta$ is given by
\begin{equation}
\beta=\frac{4\pi W_sR}{4\pi
R^2}=\sqrt{\frac{2r_s}{R}}.\label{betapolymer}
\end{equation}
And the capacitance is now
\begin{equation}
C=\frac{DRL}{2r_s}=\frac{DR^2}{r_s}N_i.\label{Cpolymer}
\end{equation}
Using Eqs.~(\ref{Ncdscreen}) and (\ref{scdscreen}), we have
\begin{eqnarray}
N_{c,d}&=&N_i\left[1\mp\sqrt{\frac{\pi
b\alpha}{\sqrt{2}}}\frac{R\eta}{Ze}\left(\frac{R}{r_s}\right)^{\frac{3}{4}}\right],
\label{Ncdstripe}\\
s_{c,d}&=&\frac{1}{v_0}\exp\left[-\frac{r_sZ^2e^2}{2k_BTDR^2}
\left(1\pm 2\sqrt{\frac{\pi
b\alpha}{\sqrt{2}}}\frac{R\eta}{Ze}\left(\frac{R}{r_s}\right)^{\frac{3}{4}}\right)\right].
\label{scdstripe}\nonumber\\
\end{eqnarray}
The phase diagram is shown in Fig.~\ref{fig:basicphase}b.

It is easy to check that in both two cases, slopes $N_{c,d}$ are
close to $N_i$ and the approximation~(\ref{mucNNi}) is valid.
Comparing $s_{c,d}$ and $N_{c,d}$ with results for spherical and
cylindrical macroions, we see that aggregation domain of the
artificial chromatin is the smallest of all three cases if
parameter $R\eta/Ze$ is small (the polymer makes many turns around
the sphere). The main reason is that in artificial cromatin,
$\epsilon$ and $\mu_c$ originate from two different kinds of
correlations. At $R\eta\ll Ze$, the correlation energy of two
polymer turns of the touching sphere is much smaller than the
correlation energy of the sphere in the complex (it interacts with
all turns of the polymer).

An artificial chromatin system in the presence of monovalent salt
is studied in experiments\cite{Sivan}, where the phase diagram of
reentrant condensation is found and the domain of partial
aggregation is not observed. This is qualitatively consistent with
our theory that the relative size of the partial aggregation
domain is small due to screening effect of monovalent salt.

\begin{figure}[ht]
\begin{center}
\includegraphics[width=0.3\textwidth]{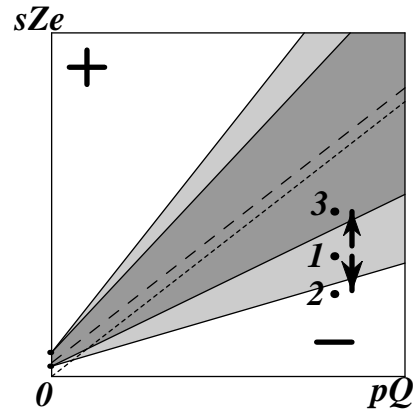}
\end{center}
\caption{The chromatin regulation and compaction. $sZe$ is the
concentration of charge of histones, $pQ$ is the concentration of
DNA. All lines and charges have the same meaning as in
Fig.~\ref{fig:basicphase}. Normal chromatin is at the point
\textbf{1}. Acetylation of histones, moves it to the point
\textbf{2} (downward arrow). On the other hand, addition of
protamines during spermatogenesis compacts chromatin into the more
condensed state \textbf{3} (upward arrow).} \label{fig:phaseregu}
\end{figure}

The phase diagram Fig.~\ref{fig:basicphase}b may be useful for
qualitative understanding of chromatin regulation. It is
known\cite{cell} that chromatin can switch from the compact state
of $30$ nm fiber or higher order structures to the loose state of
10 nm fiber in which gene transcription takes place. This
transition is caused by acetylation of core histones, which
reduces the number of positive charges of histone octamers.
Experiments in vitro\cite{Wolffe} show that the transition happens
when the charge of the octamer $Ze$ (about $+160e$ in normal
conditions) is reduced by $+12e$, i.e., only by $8\%$. Such a
sensitive response can be understood in our theory as shown in the
phase diagram Fig.~\ref{fig:phaseregu} in the plane of total
charges of octamers and DNA. Normal chromatin seems to be
partially aggregated and the system is at state $1$. Under
acetylation, the charge of octamers is reduced so that the system
moves down to state $2$ below the partial aggregation domain. The
small partial aggregation domain is crucial for the sensitive
response. Fig.~\ref{fig:phaseregu} also helps to understand
spermatogenesis in which normal chromatin is compacted further by
adding strongly positively charged proteins,
protamines\cite{cell}. In the solution, the concentration of the
positive charge increases and the system moves up from state $1$
to state $3$ in the total aggregation domain.

To conclude this section, let us make a comment about the role of
Manning condensation of monovalent salt on DNA. Above we literally
dealt with the case when the linear charge density of the polymer
$\eta$ is smaller than critical density $\eta_c = k_BT/e$ of
Manning condensation. As is well known, for a free polymer with
$\eta>\eta_c$ (for example, for DNA double helix, which has
$\eta=4.2\eta_c$), $\eta$ is renormalized to $\eta_c$ . For DNA
wrapping a positive sphere in artificial chromatin, this
renormalization is different because of the surface charge of
positive spheres. Such renormalization and its consequence on the
phase diagram were studied in Ref.~\onlinecite{Nguyen-reentrant}.

%%%%%%%%%%%%%%%%%%%%%%%%%%%%%%%%%%%%%%%%%%%%%%%%%%%%%%%%%%%%%
\section{Conclusion}\label{sec:conclusion}
%%%%%%%%%%%%%%%%%%%%%%%%%%%%%%%%%%%%%%%%%%%%%%%%%%%%%%%%%%%%%
In this paper we developed a phenomenological theory of
complexation and aggregation in asymmetric water solutions of
strongly charged negative and smaller positive colloids (macroions
and $Z$-ions) in the presence of a large concentration of
monovalent salt. We showed that in contrary with earlier
theory~\cite{Nguyen-reentrant} mostly devised for salt free
solutions, aggregate of complexes can carry net charge of
macroions and $Z$-ions which is almost as large as the charge of
free complexes. Only small touching regions of each aggregated
complex are depleted of the net charge. As a result screening by
monovalent salt leads to broader aggregation domain in the phase
diagram of the solution and to narrower, but still finite domain
of partial aggregation where aggregate is in equilibrium with free
complexes. Our phenomenological theory expressed properties of the
phase diagram through several parameters which depend on charges
of macroions and $Z$-ions, their sizes, shapes, flexibility and
screening of monovalent salt. For three different pairs of
macroions and $Z$-ions we evaluated these parameters and discussed
differences and trends in their phase diagrams. We found out that
one of these systems, which we call artificial chromatin (see
Fig.~\ref{fig:necklace}), indeed has phase diagram qualitatively
similar to the one of natural chromatin.

\begin{acknowledgments}
The authors are grateful to S. Grigoryev, A. Yu. Grosberg, V.
Lobaskin, T.T. Nguyen, and E. Raspaud for useful discussions. This
work is supported by NSF No. DMR-9985785 and DMI-0210844.
\end{acknowledgments}
%%%%%%%%%%%%%%%%%%%%%%%%%%%%%%%%%%%%%%%%%%

%%%%%%%%%%%%%%%%%%%%%%%%%%%%


\begin{thebibliography}{100}
\bibitem{Rayleigh note}This behavior is
different from the one resulting from the assumption that for a
given ($p$, $s$) all macroion-$Z$-ions complexes have same charge.
Such assumption would lead to a picture of aggregate as an
uniformly charged liquid of complexes. According to the classical
Reyleigh theory\cite{Rayleigh} of instability of a charged liquid,
this liquid would form droplets with the size optimizing the sum
of their surface and charging energy (see also
Ref.~\onlinecite{DRO} and references therein). Possibility to
redistribute $Z$-ions between complexes makes the system more
flexible and leads to a deeper minimum of the total free energy.
\bibitem{notation}The definition of $\epsilon$ is
different from Ref.\onlinecite{Nguyen-reentrant}. There notation
$\varepsilon$ is used to define the additional correlation energy
per $Z$-ion in the aggregate. The definition used here is more
convenient for systems shown Fig.~\ref{fig:sphmacroion} and
Fig.~\ref{fig:cylinmacroion} in which not all $Z$-ions in the
aggregate can gain the additional correlation energy.
\end{thebibliography}
\end{document}